\documentclass[
aps,%
12pt,%
final,%
notitlepage,%
oneside,%
onecolumn,%
nobibnotes,%
nofootinbib,
superscriptaddress,%
noshowpacs,%
centertags]%
{revtex4-1}
\RequirePackage{graphicx}

\def\refitem#1{\relax}

\begin{document}
\title{Chiral and angular momentum content of mesons}

\author{\firstname{L. Ya.} \surname{Glozman}}
\email{leonid.glozman@uni-graz.at}
\affiliation{Institute for  Physics, Theoretical Physics Branch,
 University of Graz, 
A-8010, Graz, Austria}

\begin{abstract}
First, we overview the present status of the effective chiral
restoration in excited hadrons and an alternative explanation
of the symmetry observed in the highly excited hadrons. Then we
discuss a method how to define and measure in a gauge invariant manner
the chiral and angular momentum content of mesons at different
resolution scales, including the infrared scale, where  mass
is generated. We illustrate this method by presenting results on
chiral and angular momentum content of $\rho$ and $\rho'$ mesons obtained
in dynamical lattice simulations. The chiral symmetry is strongly
broken in the $\rho(770)$ and neither the $a_1(1260)$ nor the $h_1(1170)$
can be considered as its chiral partners. Its angular momentum content
in the infrared is approximately the $^3S_1$ partial wave, in agreement
with the quark model language. However, in its first excitation, $\rho(1450)$,
the chiral symmetry breaking is much weaker and in
the infrared this state belongs predominantly to the (1/2,1/2) chiral
representation. This state is dominated  in the infrared 
by the $^3D_1$ partial wave and
cannot be considered as the first radial excitation of the $\rho(770)$-meson,
in contrast to the quark model.
\end{abstract}

\maketitle

\section{Parity doubling and higher symmetry seen in highly excited hadrons}

The spectra of highly excited hadrons, both baryons \cite{G1} and mesons \cite{G2},
reveal almost systematical parity doubling. This parity doubling can be
interpreted as an indication of effective chiral and $U(1)_A$ restorations, 
for reviews see \cite{G3}. The effective chiral restoration means that
dynamics of chiral symmetry breaking in the vacuum is almost irrelevant
to the mass generation of these highly excited hadrons and their mass
comes mostly from the chiral invariant dynamics. This is just in contrast
to the lowest lying hadrons such as $\pi$, $\rho$ or $N$, where the
chiral symmetry breaking in the vacuum is of primary importance for their
mass origin. The latter can be seen from the SVZ sum rules \cite{Shifman,Ioffe}
and many different microscopical models.

However, there could be other reasons for parity doubling \cite{JPS,SV,K,A}
and one needs alternative evidences. If the effective chiral restoration
is correct, then the highly excited hadrons should have small diagonal
axial coupling constants. It is not possible, unfortunately, to measure these
quantities experimentally. The effective chiral restoration also predicts
that the states with almost restored chiral symmetry should have small
decay coupling constants into the ground state and the pion. The decay
coupling constants can be obtained from the known decay widths. It turns out
that  all excited nucleons that have an approximate chiral partner
have a very small decay coupling constant $g_{N^*N\pi}$ (as compared
to the pion-nucleon coupling constant). In contrast, the $3/2^-, N(1520)$
state, in which case a chiral partner cannot be identified from the
spectrum, has a decay coupling that is even larger than the pion-nucleon
coupling. One observes a 100\% correlation of the spectroscopic patterns
with the $\pi N$ decays as predicted by effective chiral restoration  
\cite{G4}.

The observed high lying spectra have higher degeneracy. The states group not
only into possible chiral multiplets, but also  states with different spins
are approximately degenerate. Chiral symmetry cannot connect states with
different spins.
This means that higher symmetry is observed,
that includes chiral $SU(2)_L \times SU(2)_R$ and $U(1)_A$ as subgroups.
It is a key question to understand this high symmetry and its dynamical origin.
The answer to this question would clarify the origin of confinement and its
interconnection with  dynamical chiral symmetry breaking, the mass and 
the angular
momentum generation in QCD. It is possible to explain this  degeneracy
of states with different spins if one assumes a principal quantum number 
$n + J$ on top of chiral and $U(1)_A$ restorations \cite{GN1}.
  
If chiral restoration is correct, then there must be chiral partners
to mesons with the highest spin states at the bands around 1.7 GeV,
2 GeV and 2.3 GeV, that are presently missing, see Fig. 2 of ref. \cite{G3}.
Consequently, a key question is whether these states do not exist or
they could not be seen due to some kinematical reasons. It turns out
that the latter is correct and a centrifugal repulsion in the $\bar p p$
incoming wave suppresses {\it all}  missing states as compared to
to {\it all} observed ones \cite{GS}. There is  a weak signal for
 missing states once a careful analysis is done. Obviously, the missing states
should be also searched in other types of experiments. The same centrifugal
suppression in the pion-nucleon
scattering is  present for {\it all} missing chiral partners in the
nucleon and delta spectra \cite{K}.

The alternative explanation of the large degeneracy seen in both
nucleon and meson spectra would be existence of the relation
$M^2 \sim n + L$, where L is the orbital angular momentum in the
state. The total angular momentum $J$ is constructed from the quark
spins $S$ and the orbital angular momentum $L$ according to the standard
nonrelativistic rules. The parity of the state is connected with 
$L$ by the standard
nonrelativistic relation \cite{SV,K,A}. In such case the parity doubling is 
accidental and is not related with  chiral symmetry in the 
states. This scenario
requires that there must not be parity partners to the highest spin
states in every band. Such relation  implies that there are 
three independent
conserved angular momenta, $L,S,J$. If the high lying states 
behaved non-relativistically and assuming absence of the spin-orbit force,
it would be indeed possible to obtain a principal quantum number $ n + L$,
like in the nonrelativistic Hydrogen atom.

Such a scenario is inconsistent with QCD and can be ruled out
on very general grounds. (i) In QCD, that is a highly relativistic
quantum field theory, there is only one conserved angular momentum, $J$.
There are no representations of the Poincar\'{e} group that would contain
the orbital angular momentum $L$ as a good quantum number.
(ii) QCD is a renormalizable quantum field theory. The hadron mass is
a renormalization group invariant and does not depend on the renormalization
scale. At the same time $L$ is not a renormalization group invariant. Then
the relation $M^2 \sim  L$ cannot exist within QCD. 

From the theoretical side, there exists a transparent model that
manifestly exhibits effective chiral restoration in hadrons with
large $J$ \cite{WG,NB}. While this model is a simplification of
QCD, it gives the insight into phenomenon. The model is confining,
chirally symmetric and provides dynamical breaking of chiral symmetry
in the vacuum \cite{Y,ADLER}. The chiral symmetry breaking 
is important only at small momenta of quarks.
But at large $J$ the centrifugal
repulsion cuts off the low-momenta components in  hadrons  
 and consequently the hadron wave function and its mass are
insensitive to the chiral symmetry breaking in the vacuum. The
chiral symmetry breaking in the vacuum represents only a tiny
perturbation effect: Practically the whole hadron mass comes
from the chiral invariant dynamics.

\section{The chiral content of mesons from first principles}

To resolve the issue one needs direct information about the
chiral structure of states, which can be obtained from ab initio
lattice simulations. Here we  define and reconstruct
in dynamical lattice simulations a chiral as well as an angular
momentum decomposition of the leading quark-antiquark component
of mesons \cite{GLL1,GLL2}. 

The variational method \cite{var} represents a tool to study
the hadron wave function. One chooses a set of interpolators
$\{O_1,O_2,\ldots,O_N\}$ with the same quantum numbers as the
state of interest and computes the cross-correlation matrix 
$$
C_{ij}(t) = \big\langle\; O_i(t)\; O_j^\dag (0)\;\big\rangle\ .
$$

\noindent
If this set   is complete and orthogonal with
respect to some transformation group, then it allows to define
a content of a hadron in terms of representations of this
group.

In \cite{G2,G3,CJ} a classification of
all non-exotic
quark-antiquark states (interpolators) in the light meson sector
 with
respect to the $SU(2)_L\times SU(2)_R$ and $U(1)_A$ was done. 
If no explicit excitation of the
gluonic field with the non-vacuum quantum numbers is present, this basis is a
complete one for a quark-antiquark system and we can define and
investigate chiral symmetry
breaking in a state.  The eigenvectors of the cross-correlation matrix
describe the
quark-antiquark component of the state in terms of different chiral representations. 

For example, when we study the $\rho$ meson and its excitations, two
different chiral representations exist that are consistent with the quantum
numbers of the $\rho$-mesons. 
Assume that
chiral symmetry is not broken. Then there are two independent states.
The first one is $\vert(0,1)\oplus (1,0);\, 1\, 1^{--}\rangle$;
it can be created from the vacuum by the standard
vector current, 
$
O_V = {\bar q} \gamma^i \vec{\tau} q. 
$
Its chiral partner is the $a_1$ meson.
 The other state is $\vert(1/2,1/2)_b;\, 1\, 1^{--}\rangle$, which
 can be created by the pseudotensor operator,
$
O_T = {\bar q} \sigma^{0i} \vec{\tau} q. $
The chiral partner is the $h_1$ meson.
 
Chiral symmetry breaking in a state implies that the state should
 in reality be a mixture of both representations. If the state is
 a superposition of both representations with approximately equal
 weights, then the chiral symmetry is maximally violated in the state.
 If, on the contrary, one of the representations strongly dominates
 over the other representation, one could speak about effective chiral 
 restoration
 in this state.
 
 Diagonalizing the cross-correlation matrix one can extract energies
 of subsequent states from the leading exponential decay of each eigenvalue
 
 $$
C_{ij}(t) = \big\langle\; O_i(t)\; O_j^\dag (0)\;\big\rangle\
          = \sum_n\, a_i^{(n)} a_j^{(n)*} e^{-E_n t}\ . 
$$
The corresponding eigenvectors give us information about the
structure of each state. Namely,
the coefficients $a_i^{(n)}$ define the overlap of the physical state
$\vert n\rangle$ with the  interpolator $O_i$,
$
a_i^{(n)} = \langle 0 \vert O_i \vert n \rangle\ .
$

While the absolute value of the coupling constant $a_i^{(n)}$
cannot be defined in lattice simulations (because a 
normalization of the quark fields on the lattice is arbitrary),
their ratio for two different operators $O_i$ and for a given
state is well defined \cite{GLL1}.
 Consequently, 
the ratio of the vector to pseudotensor couplings, $a_V^{(n)}/a_T^{(n)}$, tells us
about the chiral symmetry breaking in the states $n=\rho,\rho'$.

\section{The angular momentum content of mesons from first principles}

The chiral representations can be transferred into the 
${}^{2S+1}L_J$ basis, using the
unitary transformation \cite{GN1,GN2}
\begin{equation}
\left( \begin{array}{c}
\vert(0,1)\oplus (1,0);\, 1\, 1^{--}\rangle\\
\vert(1/2,1/2)_b;\, 1\, 1^{--}\rangle\\
\end{array} \right)
=
U\cdot
\left( \begin{array}{c}
\vert 1;\, {}^3S_1\rangle\\
\vert 1;\, {}^3D_1\rangle\\
\end{array} \right)\ ,
\end{equation}
where $U$ is given by
\begin{equation}
U=\left( \begin{array}{lr}
\sqrt{\frac{2}{3}}\ &\  \sqrt{\frac{1}{3}}\\
\sqrt{\frac{1}{3}}\ &\ -\sqrt{\frac{2}{3}}\\
\end{array} \right)\ .
\end{equation}

\noindent
Thus, using the interpolators $O_V$ and $O_T$  
for  diagonalization of the cross-correlation matrix, we are
able also to reconstruct a partial wave content of the leading $\bar q q$
Fock component of
the $\rho$-mesons. Note, that it is a manifestly gauge-invariant
definition of the angular momentum content of mesons.

\section{Scale dependence of the chiral and angular momentum decompositions}

The ratio $a_V^{(n)}/a_T^{(n)}$ as well as a partial wave content of
a hadron are not the renormalization group invariant quantities. Hence
they manifestly depend on a resolution scale at which we probe a
hadron.
If we probe the hadron structure with the local interpolators,
then we study the hadron decomposition at the scale fixed by
the lattice spacing $a$. For a reasonably small $a$ this scale
is close to the ultraviolet scale. However, we are interested 
in the hadron content at the infrared scales, where mass is
generated. For this purpose we cannot use a large $a$, because
matching with the continuum QCD will be lost. Given a fixed,
reasonably small lattice spacing $a$ a small resolution scale $1/R$
can be achieved by the gauge-invariant smearing of the point-like
interpolators. We smear every quark field in spatial directions 
with the Gaussian profile over the size $R$ in
physical units such that $R/a\gg 1$, see Fig. 1. Then even in the continuum
limit $a \rightarrow 0$ we probe the hadron content at the resolution
scale fixed by $R$. Such  definition of the resolution is similar
to the experimental one, where an external probe is sensitive only to
quark fields (it is blind to gluonic fields) at a resolution 
that is determined by the momentum transfer in spatial directions.

\begin{figure}[tb]
 \includegraphics[height=.3\textwidth]{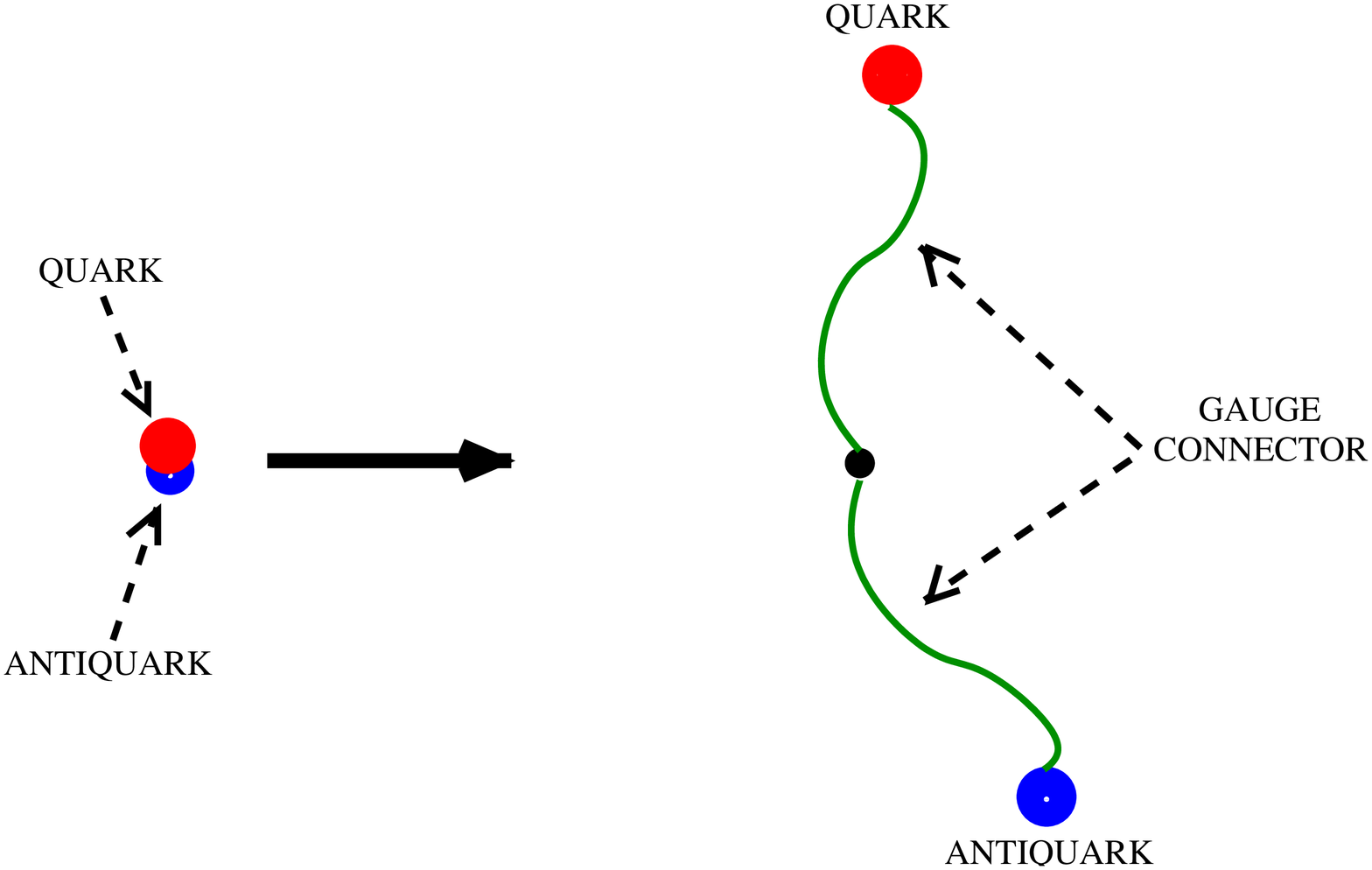}\hfill
 \includegraphics[height=.3\textwidth]{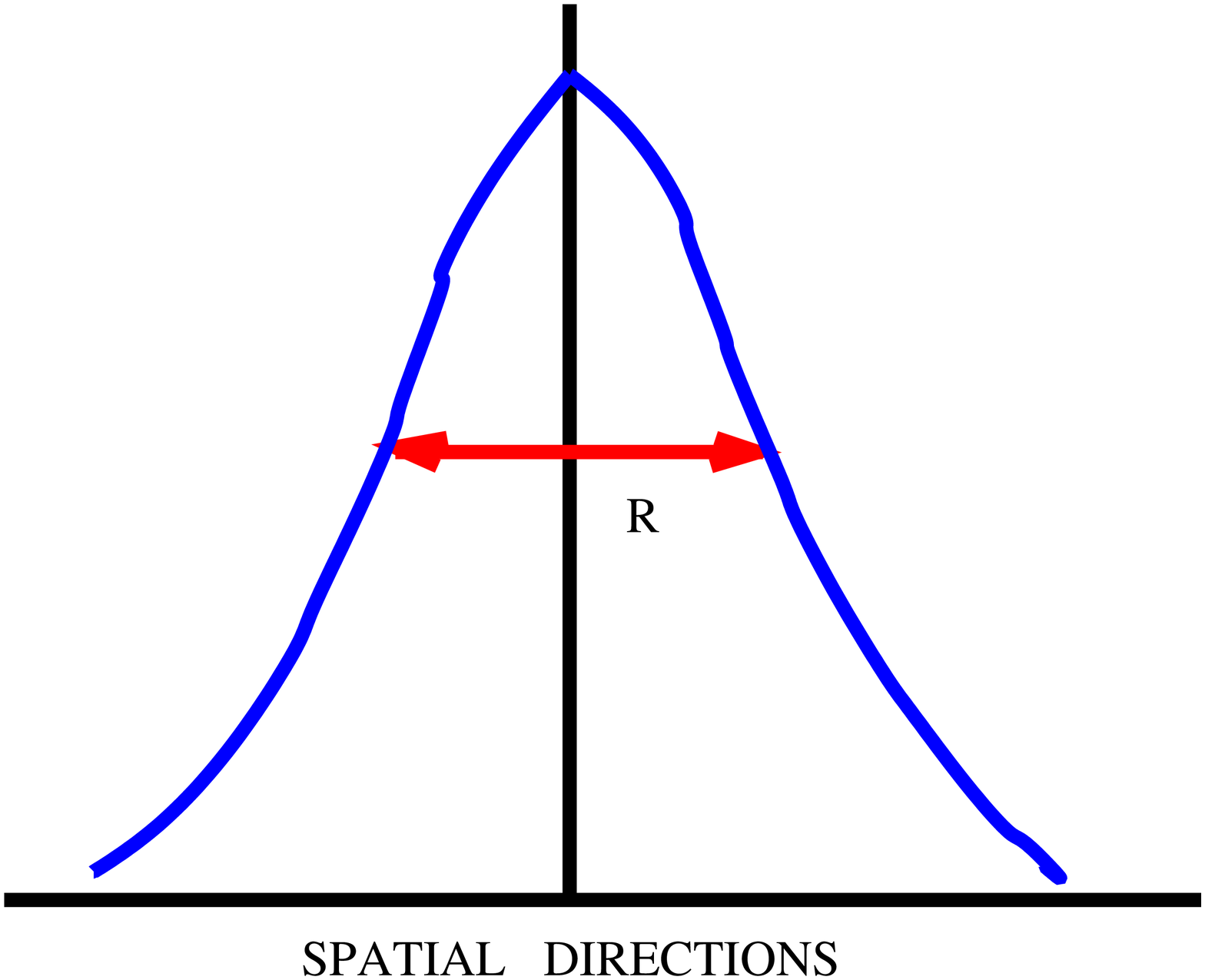}
 \caption{Gauge-invariant smearing and the resolution scale $R$ definition}
\end{figure}

\section{The chiral and angular momentum content of $\rho$ and $\rho'$ mesons}

To explore the chiral structure of mesons and  possible
effective chiral restoration it is important to have a Dirac operator with
good chiral properties. We use specifically the Chirally Improved Dirac
operator \cite{GAT}. The set of dynamical configurations is used for
two mass-degenerate light sea quarks, see for details ref. \cite{GLL2}.

Our cross-correlation matrix is calculated with the following 
four interpolators

$$
O_1=\bar u_n \gamma^i d_n,~
O_2=\bar u_w \gamma^i d_w, ~
O_3=\bar u_n \gamma^t \gamma^i  d_n, ~
O_4=\bar u_w \gamma^t \gamma^i  d_w,
$$

\noindent
where$\gamma^i$ is one of the spatial Dirac matrices, $\gamma^t$ is the
$\gamma$-matrix in (Euclidean) time direction. The subscripts $n$ and $w$ (for
narrow and wide) denote the two smearing widths, $R\approx 0.34$ fm  and $0.67$
fm, respectively. Both the ground
state mass and the mass of the first excited state of the $\rho$-meson are
shown on the l.h.s.\ of Fig.~2.

\begin{figure}[tb]
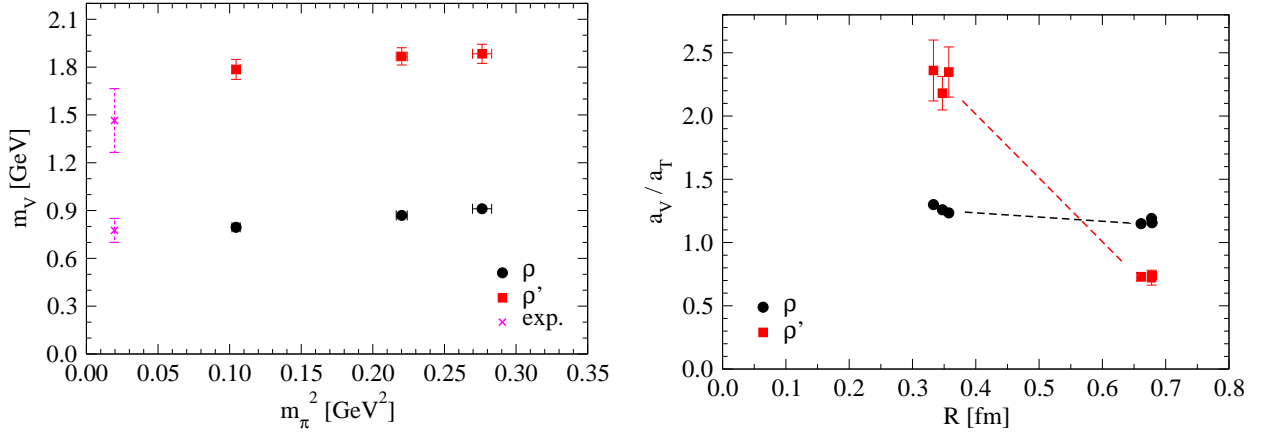

\includegraphics[width=80mm,clip]{mass_rho.eps}\hfill
\includegraphics[width=80mm,clip]{ratio_vs_R.eps}
\caption{L.h.s.: The vector meson mass $m_V$ is plotted against $m_\pi^2$. 
 Black circles represent the ground state, $\rho$, and red
squares represent the first excitation, $\rho'$. The experimental values are
depicted as magenta crosses with decay width indicated. R.h.s.: The ratio
$a_V/a_T$ is plotted against the smearing width $R$ for all three pion masses.
Black circles represent the ground state and red squares the first excitation.
Broken lines are drawn only to guide the eye.}
\end{figure}

On the r.h.s.\ of Fig. ~2 we show the $R$-dependence of the ratio  
$a_V/a_T$ both for the ground state
$\rho$-meson and its first excited state.  For the ground state at
the smallest resolution scale of $R\approx 0.67$ fm this ratio is approximately
$1.2$, i.~e., we see a strong mixture of the
$(0,1)\oplus (1,0)$ and $(1/2,1/2)_b$ representations in the $\rho$-meson. 
Consequently, there is no chiral partner to $\rho(770)$.
Such ratio implies that the vector meson in the infrared is
approximately a ${}^3S_1$ state with a tiny admixture of a ${}^3D_1$ wave.

However, the situation changes dramatically for the first excited state, $\rho'
= \rho(1450)$. In this case a strong dependence of the ratio on the resolution
scale is observed.
Although we do not have the precise value of the ratio $a_V/a_T$ for
$\rho(1450)$ at large $R \sim 0.8 - 1$fm, 
it is indicative that this value is very small.
One observes a significant contribution  from the $(1/2,1/2)_b$
representation and a contribution of the other representation is
suppressed. This indicates a smooth onset of effective chiral restoration.
The approximate chiral partner is $h_1(1380)$.
This small ratio also implies
a leading contribution of the ${}^3D_1$ wave. 
 This
result is inconsistent with $\rho'$ to be a radial excitation of the ground
state $\rho$-meson, i.~e., a ${}^3S_1$ state, as predicted by the quark model.

\medskip
\begin{center}
{\bf Acknowledgements}
\end{center}

The author is thankful to Christian Lang and Markus Limmer for
a fruitful collaboration on the lattice aspects of this talk.
Support of the Austrian Science Fund through the grant P21970-N16 
is acknowledged.

\end{document}